\documentclass{article}

\usepackage[final,nonatbib]{neurips_2024}

\usepackage[framemethod=TikZ]{mdframed}
\mdfsetup{%
   middlelinewidth=2pt,
   roundcorner=10pt}

\usepackage[utf8]{inputenc} 
\usepackage[T1]{fontenc}    
\usepackage{url}            
\usepackage{booktabs}       
\usepackage{amsfonts}       
\usepackage{nicefrac}       
\usepackage{microtype}      
\usepackage{xcolor}         

\usepackage{amssymb}
\usepackage{microtype}
\usepackage{inconsolata}
\usepackage{graphicx}
\usepackage{xspace}
\usepackage{wrapfig}
\usepackage{multirow}
\usepackage{tablefootnote}
\usepackage{hyperref}       

\newcommand{\model}{RNA-GPT\xspace}
\newcommand{\dataset}{RNA-QA\xspace}
\newcommand{\datasize}{407,616\xspace}

\title{\raisebox{-0.8ex}{\includegraphics[height=28pt]{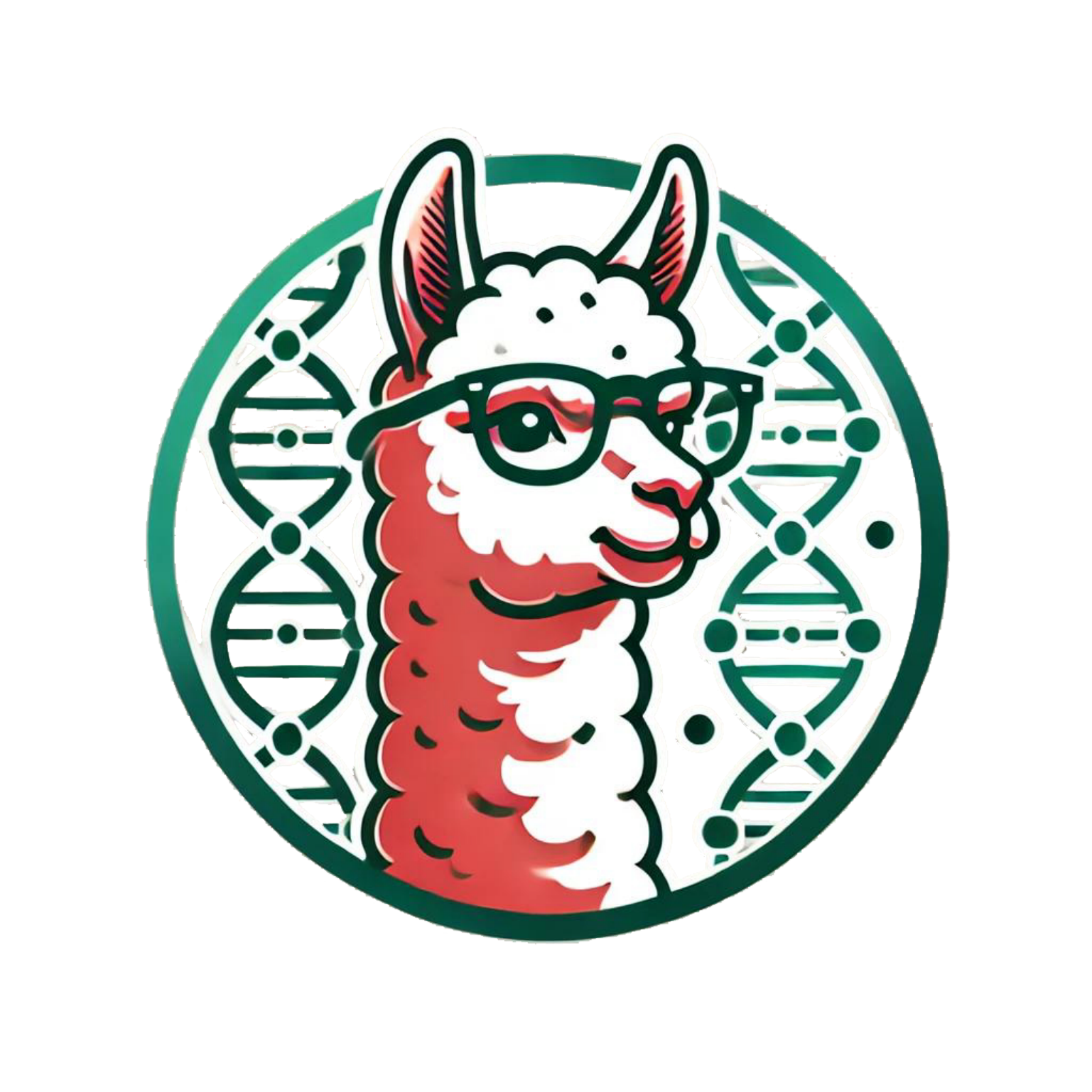}} RNA-GPT: Multimodal Generative System for RNA Sequence Understanding}

\author{
    Yijia Xiao$^{1}$, 
    Edward Sun$^{1}$, 
    Yiqiao Jin$^{2}$, 
    Wei Wang$^{1}$ \\ \\
    $^{1}$University of California, Los Angeles, \\
    $^{2}$Georgia Institute of Technology, \\
    \texttt{\{yijia.xiao,wei.wang\}@cs.ucla.edu, edwardsun12895@g.ucla.edu,} \\
    \texttt{yjin328@gatech.edu}
}

\begin{document}

\maketitle

\begin{abstract}
RNAs are vital molecules that carry genetic information essential for life, with significant implications for drug development and biotechnology. However, RNA research is often slowed by the vast amount of literature. To address this, we introduce \model, a multi-modal RNA chat model that simplifies RNA discovery by leveraging extensive RNA literature. \model combines RNA sequence encoders with linear projection layers and state-of-the-art large language models (LLMs) for precise representation alignment. This enables it to process user-uploaded RNA sequences and provide concise, accurate responses. Our scalable training pipeline, powered by \dataset, automatically gathers RNA annotations from RNACentral using a divide-and-conquer approach with GPT-4o and latent Dirichlet allocation (LDA) to handle large datasets and generate instruction tuning samples. Experiments show \model effectively handles complex RNA queries, streamlining RNA research. We also introduce \dataset, a \datasize RNA dataset for modality alignment and instruction tuning.
\end{abstract}

\section{Introduction}

Large language models (LLMs) trained on internet-scale corpora have been shown to perform extraordinarily well on a large array of tasks from Olympiad-level mathematical and scientific reasoning to planning long-term tasks for robotic systems ~\cite{xiao2024logicvistamultimodalllmlogical, he2024olympiadbenchchallengingbenchmarkpromoting, huang2022innermonologueembodiedreasoning}. Recent advances in the biological and medical fields have enabled the adaptation of powerful models to accelerate research, significantly reducing reliance on traditional experiments. Since proteins, RNAs, and DNAs can be represented as character strings and a vast amount of sequenced data is readily available, this has created an ideal environment for training language models to predict and generate protein, DNA, and RNA structures and sequences. Protein language models like ESM have successfully encoded protein sequence and structure information, inspiring works such as ProteinGPT and ProtSt, which adapt protein representations into a language-based format, enabling natural language querying of protein data~\cite{ESM,ESM-2,ESM-1v,elnaggar2021prottrans,xiao2021modeling, xiao2024proteingptmultimodalllmprotein, xu2023protst}. Similar to ESM-2, works like RiNALMo and RNA-FM have utilized the flexible capabilities of LLMs to learn and predict RNA structure and functions~\cite{penić2024rinalmogeneralpurposernalanguage, chen2022interpretablernafoundationmodelRNAFM}. 

Much like the motivation behind protein research, where proteins are represented as strings of characters, RNAs—with their sequences of five unique nucleotides—have also sparked interest in computational RNA and DNA research using large language models (LLMs).

While models like ProteinGPT, ProtST, ProteinChat, and ProteinCLIP, have made significant progress in aligning protein sequences and structures with textual descriptions, advancements in the DNA and RNA domains are far less advanced~\cite{xiao2024proteingptmultimodalllmprotein, guo2023proteinchat, wu2024proteinclip, xu2023protst, wang2023instructprotein, zhuo2024protllminterleavedproteinlanguagellm}. Previous efforts, such as RiNALMo and RNA-FM have mainly focused on specific tasks like promoter or enhancer prediction, and structure and function analysis~\cite{chen2022interpretablernafoundationmodelRNAFM, penić2024rinalmogeneralpurposernalanguage, Shulgina2024RNALanguage}. ChatNT is among the few models striving to bridge the gap between RNA comprehension and natural language~\cite{Richard2024ChatNT}. However, its emphasis is more on performing biological tasks as a conversational agent rather than providing deep RNA understanding and comprehensive dialogue. As a result, there is a notable gap in RNA chat models that offer in-depth knowledge. However, applying multimodal LLMs to RNA modeling presents unique challenges, especially in integrating diverse modalities such as textual descriptions, RNA sequences, and structural data.

To overcome these challenges, we propose a two-step approach to \model. First, we utilize the RNA-FM sequence encoder to embed RNA sequences, followed by aligning these sequence representations with natural language through a large, automatically curated QA dataset from RNA Central~\cite{chen2022interpretablernafoundationmodelRNAFM, RNACentral}. Secondly, to ensure our model generates concise and accurate responses, we break down RNA-QA’s abstract summaries into individual QA pairs for instruction tuning, enhancing the model’s ability to deliver clear and relevant answers. We utilize Meta AI's flagship Llama-3 8B Instruction as our backbone LLM to provide solid general language understanding~\cite{dubey2024llama3herdmodels}. More specifically, our contributions are as follows:

\begin{itemize}
    \item \textbf{Novel Framework.} \model is one of the first multi-modal RNA sequence chat models that enables deep, interactive RNA-focused conversations, significantly enhancing the understanding of RNAs for biological research. 
    \item \textbf{Large-scale Dataset and Collection Pipeline.} We introduce \dataset, a QA dataset derived from the RNA Central Dataset for modality alignment instruction tuning of RNA chat models ~\cite{RNACentral}. We also present our highly scalable collection pipeline that automates the scraping and summarizing of relevant literature on RNA. Using a divide-and-conquer summarization strategy, we ensure that research details are preserved effectively. For over \datasize RNA samples, we create QA pairs, each accompanied by a comprehensive research summary based on available literature, and between 5 and 14 QA pairs. The depth and diversity of these annotations make \dataset an excellent resource for instruction tuning.
\end{itemize}

\section{Methodology}

\begin{figure*}[htbp]
\begin{center}
\centerline{\includegraphics[width=0.9\linewidth]{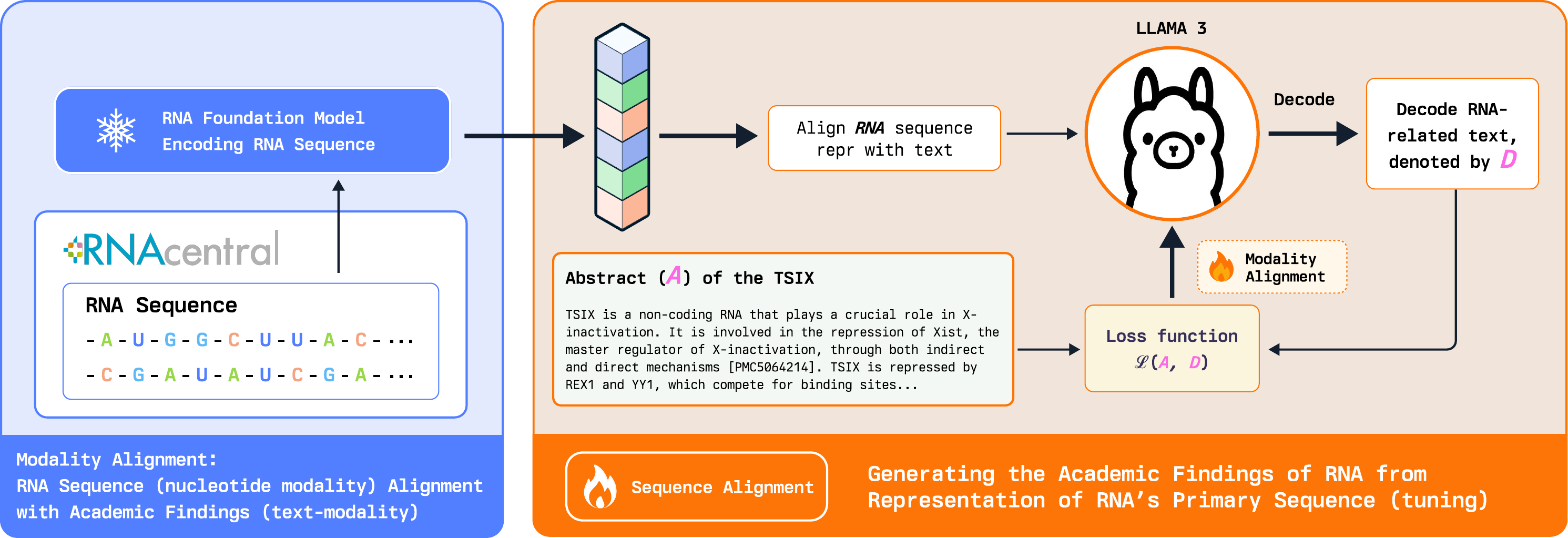}}
\caption{\model Modality Fusion \& Alignment Stage: we freeze the sequence encoder block and train the linear projection layer to learn how to align RNA sequence representations with text. In the alignment stage, the input to the training is only the projected RNA representation. No text prompts are incorporated in this stage.}
\label{fig:schema_ma}
\end{center}
\end{figure*}

\begin{figure*}[htbp]
\begin{center}
\centerline{\includegraphics[width=0.9\linewidth]{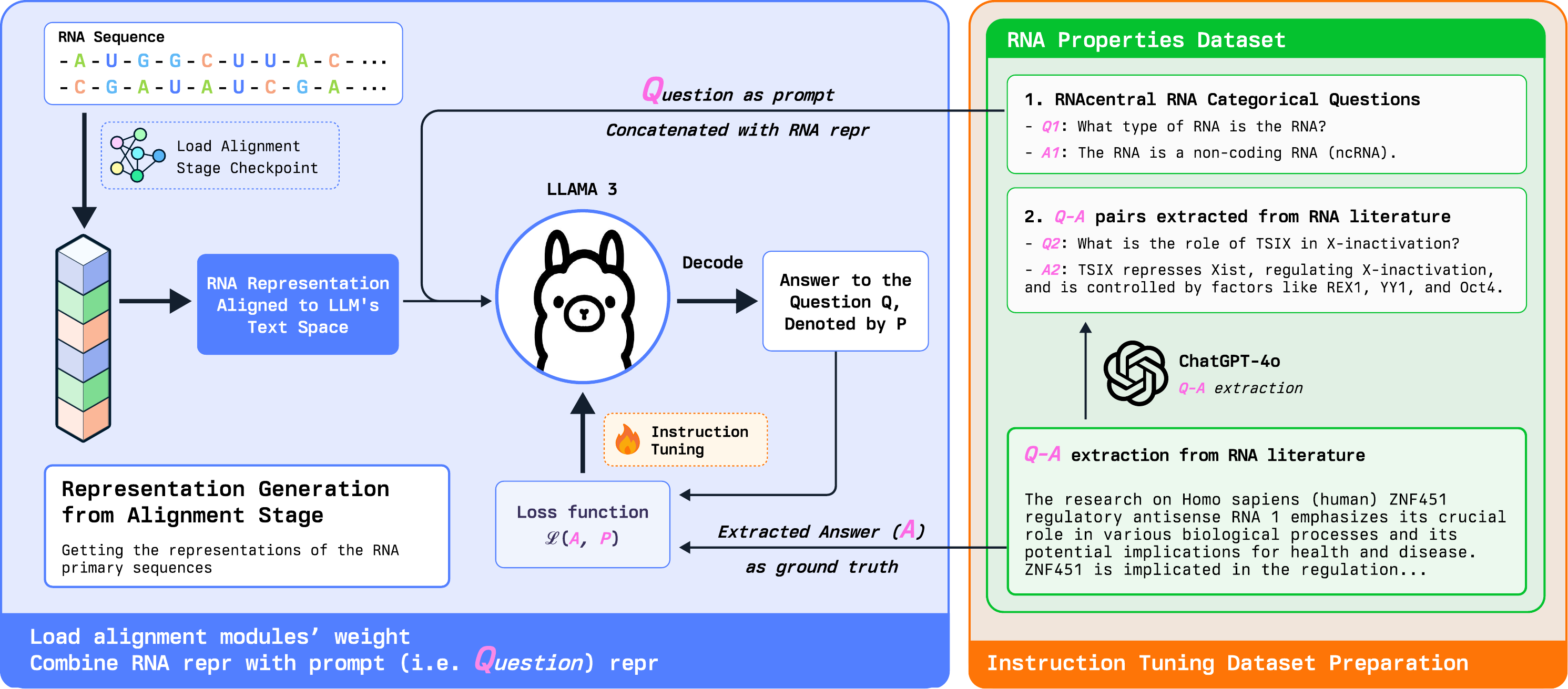}}
\caption{\model Modality Fusion \& Alignment Stage: we freeze the sequence encoder block and train the linear projection layer to learn how to align RNA sequence representations with text. In the alignment stage, the input to the training is only the projected RNA representation. No text prompts are incorporated in this stage.}
\label{fig:schema_it}
\end{center}
\end{figure*}

\subsection{Model Architecture}

\model uses the pre-trained RNA-FM sequence encoder (Figure \ref{fig:schema_ma} and Figure \ref{fig:schema_it}) to embed RNA sequences, which are then passed through a linear projection layer. This layer learns to map the RNA embeddings to a shared representation space with natural language, enabling alignment with a backbone LLM, for which we chose Meta’s Llama-3 8B model. The training process is divided into two stages: 1) {\it Sequence and Modality Alignment}, where RNA and natural language representations are aligned, and 2) {\it Instruction Tuning}, where the model is fine-tuned for task-specific QA generation.

\subsubsection*{Modality Alignment Stage (Stage 1)}
RNA sequences in the form of strings are first fed into the pre-trained sequence encoder, featuring 12 transformer layers trained with 23 million RNAs from the RNA Central database via self-supervised learning ~\cite{RNACentral, chen2022interpretablernafoundationmodelRNAFM}.
We utilize a specialized token <RNAHere> for RNA-text modality alignment: 
$$
\begin{array}{l}
    \mathbf{Q}{:}<RNA><RNAHere></RNA><QuestionPrompts>\\\mathbf{A}{:}<Description>
\end{array}   
$$

The embedded sequence information is encoded into the soft prompts and prepended to the question prompt. In stage 1 training, the question \textbf{Q} is left empty to prioritize learning the abstract description from the RNA representation. 
The description tag \textbf{A} is replaced with the full annotation from RNA Central~\cite{RNACentral} to train the linear projection layer to align an RNA with its full abstract annotation.

\subsubsection*{Instruction Tuning Stage (Stage 2)}
In stage 2, we instruction-tune the model using our curated \dataset dataset. Previous protein-related chat models rely on fully annotated abstracts, frequently resulting in excessively lengthy and irrelevant responses. We take a different approach to address this by breaking down the full annotations into targeted QA samples with concise answers to specific questions as prediction targets. This allows the chat model to provide more relevant and accurate responses.

We augment the full abstract annotation dataset from stage 1 using GPT-4o-mini to generate explicit QA pairs for this stage. The prompts from stage 1 are adapted to the Llama-3 style (``$\#\#\# Human: \ldots$'' and ``$\#\#\# Assistant: \ldots$''), with \textbf{Q} replaced by explicit questions from \dataset, such as ``What regulatory role does the RNA have along with other RNAs?'' The model then generates descriptive yet concise answers such as ``The RNA is involved in transcript splicing regulation along with RNVU1-18 and CLK1`` as \textbf{A}.

\subsection{\dataset Dataset}
\label{sec:dataset}
\begin{figure*}[htbp]
\begin{center}
\centerline{\includegraphics[width=\linewidth]{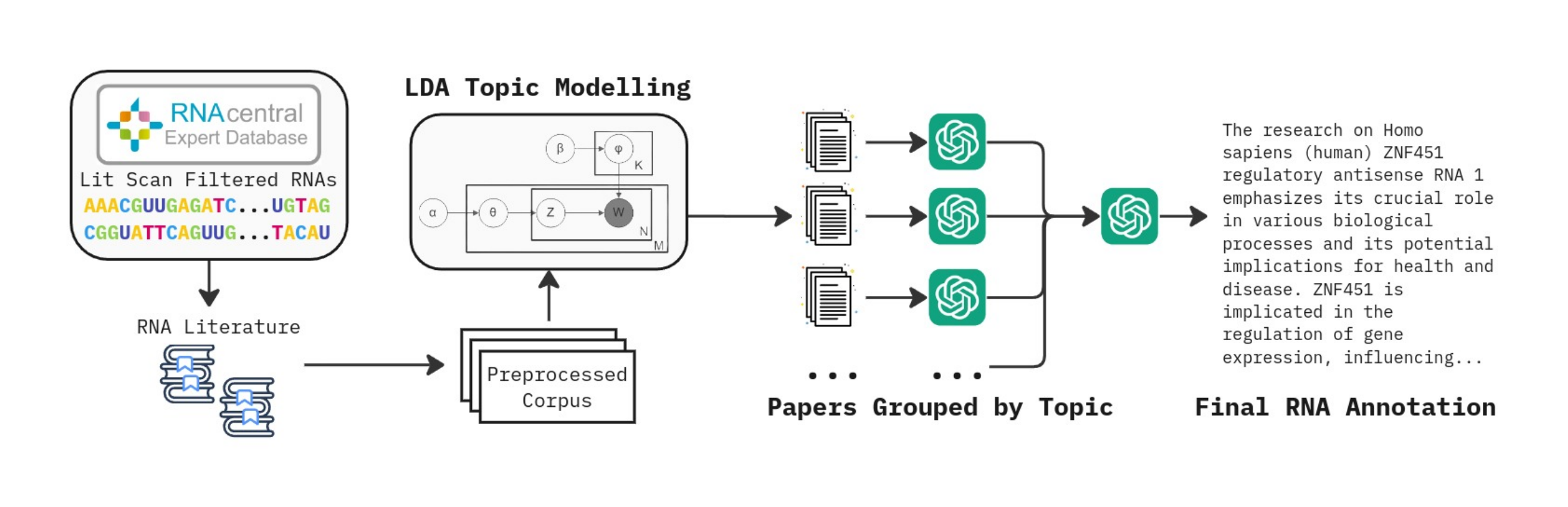}}
\caption{\dataset uses an automated pipeline to scrape and summarize existing RNA literature. We apply latent Dirichlet allocation (LDA) to group the vast literature on each RNA, and then we summarize each group individually using GPT-4o-mini. These summaries are then combined and refined to produce the final RNA annotation.}
\label{fig:dataset_gather}
\end{center}
\end{figure*}

To achieve modality alignment, we constructed a large-scale dataset from the RNA Central database~\cite{RNACentral}, comprising \datasize RNA sequences paired with abstract descriptions.

\subsubsection*{Divide and Conquer RNA Literature Summarization}
We begin by filtering RNA sequences from RNA Central~\cite{RNACentral}, focusing on those indexed with "Lit Scan," yielding around 420,000 RNAs with associated research papers. We refine this set to include only sequences up to 1024 nucleotides, the maximum input length for our sequence encoder. For the remaining \datasize RNAs, we scrape and extract abstracts from all relevant literature. As shown in Figure \ref{fig:dataset_gather}, we apply LDA topic modeling to group papers by topic, summarizing each group individually. This ensures each summarization focuses on a narrower, cohesive subject area, minimizing information loss. We have found that summarizing broad topics often causes key details to be missed, as the model struggles to condense diverse information. Grouping similar topics allows for more precise, detailed summaries that retain essential context. The final annotation is created by combining these summaries in a final round of summarization. This divide-and-conquer approach improves accuracy and efficiently handles large datasets. Moreover, it overcomes the token limits of GPT models, allowing for detailed, information-dense annotations of large RNA research profiles.

\subsection{Data Augmentation}
Similar works in protein chat alignment often use the entire protein annotation for instruction tuning~\cite{xu2023protst, guo2023proteinchat, wang2024protchatgpt, xiao2024proteingptmultimodalllmprotein}, which often result in verbose and irrelevant responses. To address this, \model decomposes the rich RNA annotations of \dataset into more specific QA-pairs for instruction tuning using GPT-4o-mini so that user instructions can be concisely answered. Concretely, we prompt GPT-4o-mini to generate both open-ended and close-ended QA pairs with the context of the \dataset annotation to decompose the abstract into atom-level QA pairs.

\section{Experiments}

We trained \model using the flagship Llama-3 8B model architecture \cite{dubey2024llama3herdmodels} using a smaller 5K RNA, 121K QA samples subset for our initial model. We are in the process of training the large \model that uses all \datasize RNAs of \dataset with millions of QA samples.

\begin{table*}[htbp]
  \centering
  \caption{RNA-QA (\textbf{AIS}): RNA Sequence (left), Modality Fusion (middle), and RNA-GPT (right). Embedding base models are BERT, PubMedBERT, and OpenAI's GPT text-embedding-3-large.}
  \resizebox{\linewidth}{!}{
  \begin{tabular}{l|rrr|rrr|rrr}
  \toprule
  & \multicolumn{3}{c|}{RNA Sequence} & \multicolumn{3}{c|}{Modality Fusion} & \multicolumn{3}{c}{RNA-GPT} \\
  Metric & $S_{\mathrm{BERT}}$ & $S_{\mathrm{Pub}}$ & $S_{\mathrm{GPT}}$ & $S_{\mathrm{BERT}}$ & $S_{\mathrm{Pub}}$ & $S_{\mathrm{GPT}}$ & $S_{\mathrm{BERT}}$ & $S_{\mathrm{Pub}}$ & $S_{\mathrm{GPT}}$ \\
  \midrule
  \textbf{Precision}  
  & 0.7372 & 0.5528 & 0.5219 & 0.6929 & 0.6507 & 0.6655 & 0.8602 & 0.7384 & 0.7848 \\
  \textbf{Recall}  
  & 0.7496 & 0.5270 & 0.5474 & 0.8028 & 0.6082 & 0.6603 & 0.8404 & 0.7208 & 0.7561 \\
  \textbf{F1 Score}  
  & 0.7424 & 0.5387 & 0.5339 & 0.7403 & 0.6283 & 0.6627 & 0.8494 & 0.7293 & 0.7700 \\
  \bottomrule
  \end{tabular}
  }
  \label{tab:scores2}
\end{table*}

We conducted a series of experiments to assess \model's effectiveness both quantitatively and qualitatively along with ablation studies to ascertain the importance of various modules at different stages. These included the original model (LLM with RNA sequence as text input), the modality-aligned model, and the final instruction-tuned model.

\begin{table*}[htbp]
  \centering
  \caption{RNA-QA (\textbf{AIS}): ROUGE Score with RNA Sequence, Modality Fusion, and RNA-GPT.}
  \resizebox{\linewidth}{!}{
  \begin{tabular}{l|rrr|rrr|rrr}
  \toprule
  & \multicolumn{3}{c|}{RNA Sequence} & \multicolumn{3}{c|}{Modality Fusion} & \multicolumn{3}{c}{RNA-GPT} \\
  Metric & Rouge-1 & Rouge-2 & Rouge-L & Rouge-1 & Rouge-2 & Rouge-L & Rouge-1 & Rouge-2 & Rouge-L \\
  \midrule
  \textbf{ROUGE}  
  & 0.2364 & 0.0935 & 0.2037 & 0.2239 & 0.1364 & 0.2091 & 0.5031 & 0.3667 & 0.4747 \\
  \bottomrule
  \end{tabular}
  }
  \label{tab:rouge2}
\end{table*}

\begin{figure}[h!]
    \centering
    \begin{minipage}{0.48\textwidth}
        \centering
        \includegraphics[width=\linewidth]{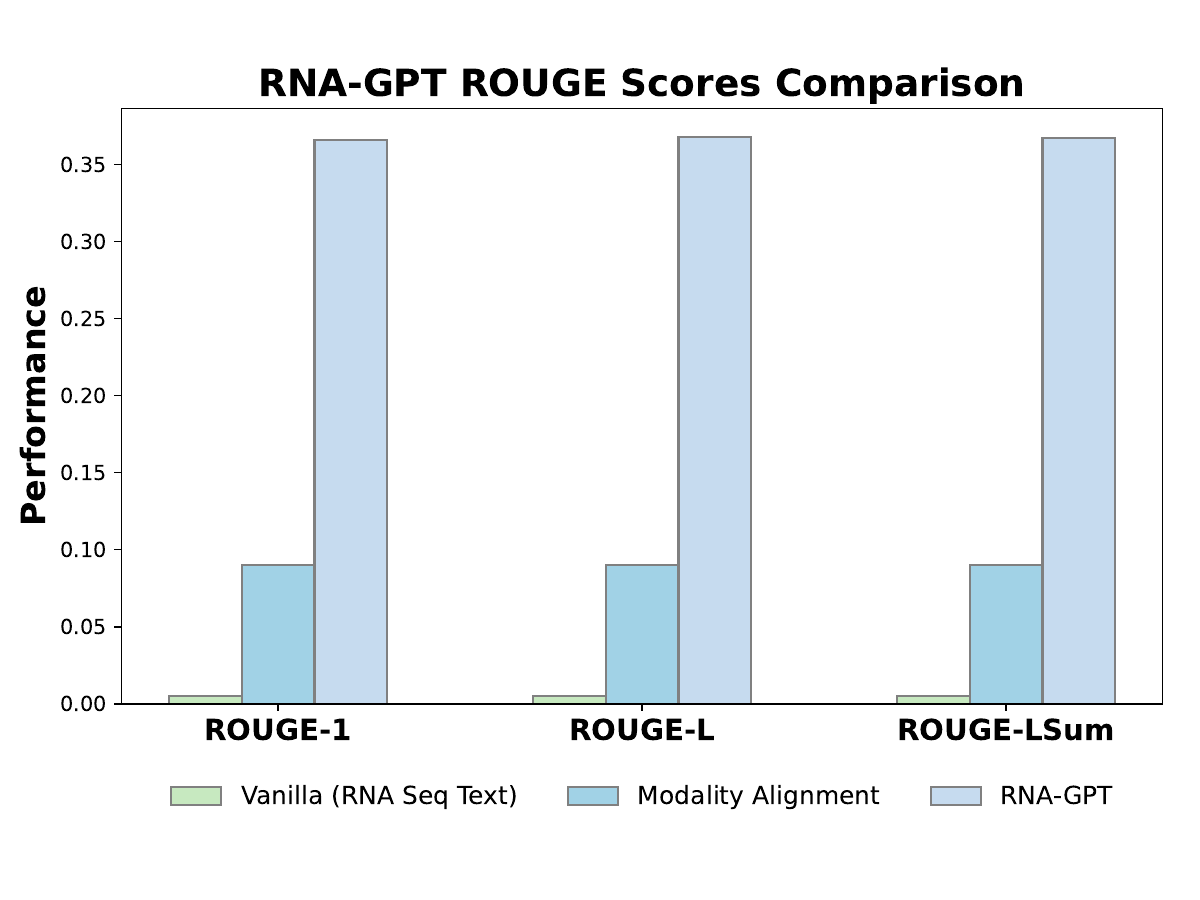}
        \caption{ROUGE Score Comparison}
        \label{fig:rouge}
    \end{minipage}
    \hfill
    \begin{minipage}{0.48\textwidth}
        \centering
        \includegraphics[width=\linewidth]{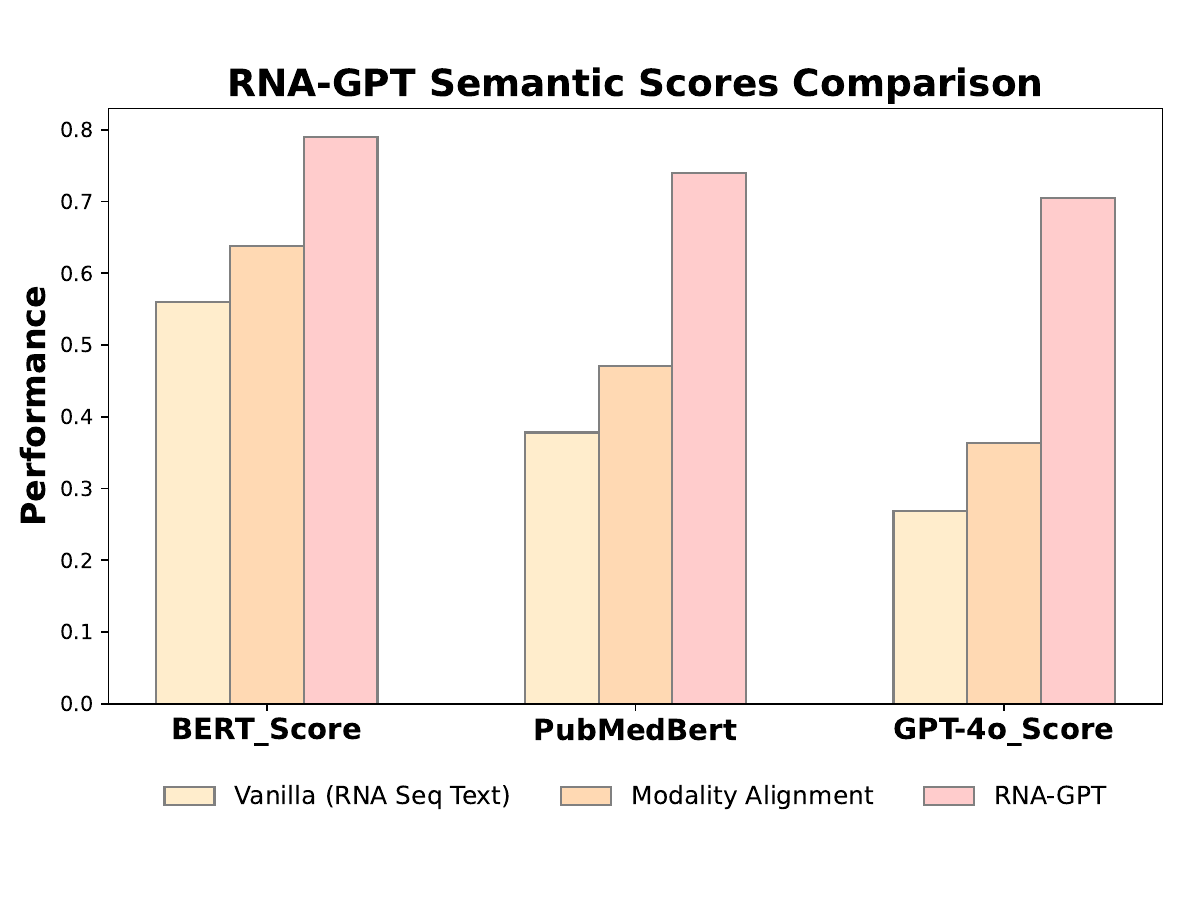}
        \caption{Semantic Score Comparison}
        \label{fig:semantic}
    \end{minipage}
\end{figure}


\begin{table*}[h!]
  \centering
  \caption{RNA-QA (\textbf{D\&C}): RNA Sequence (left), Modality Fusion (middle), and RNA-GPT (right). Embedding base models are BERT, PubMedBERT, and OpenAI's GPT text-embedding-3-large.}
  \resizebox{\linewidth}{!}{
  \begin{tabular}{l|rrr|rrr|rrr}
  \toprule
  & \multicolumn{3}{c|}{RNA Sequence} & \multicolumn{3}{c|}{Modality Fusion} & \multicolumn{3}{c}{RNA-GPT} \\
  Metric & $S_{\mathrm{BERT}}$ & $S_{\mathrm{Pub}}$ & $S_{\mathrm{GPT}}$ & $S_{\mathrm{BERT}}$ & $S_{\mathrm{Pub}}$ & $S_{\mathrm{GPT}}$ & $S_{\mathrm{BERT}}$ & $S_{\mathrm{Pub}}$ & $S_{\mathrm{GPT}}$ \\
  \midrule
  \textbf{Precision}  
  & 0.7612 & 0.5498 & 0.5479 & 0.6884 & 0.6201 & 0.6676 & 0.8620 & 0.7173 & 0.7568 \\
  \textbf{Recall}  
  & 0.7654 & 0.5512 & 0.5649 & 0.8187 & 0.5830 & 0.6602 & 0.8623 & 0.7161 & 0.7554 \\
  \textbf{F1 Score}  
  & 0.7625 & 0.5501 & 0.5561 & 0.7466 & 0.6005 & 0.6637 & 0.8609 & 0.7165 & 0.7560 \\
  \bottomrule
  \end{tabular}
  }
  \label{tab:scores_dnc}
\end{table*}

\begin{table*}[h!]
  \centering
  \caption{RNA-QA (\textbf{D\&C}): ROUGE Score with RNA Sequence, Modality Fusion, and RNA-GPT.}
  \resizebox{\linewidth}{!}{
  \begin{tabular}{l|rrr|rrr|rrr}
  \toprule
  & \multicolumn{3}{c|}{RNA Sequence} & \multicolumn{3}{c|}{Modality Fusion} & \multicolumn{3}{c}{RNA-GPT} \\
  Metric & Rouge-1 & Rouge-2 & Rouge-L & Rouge-1 & Rouge-2 & Rouge-L & Rouge-1 & Rouge-2 & Rouge-L \\
  \midrule
  \textbf{ROUGE}  
  & 0.2472 & 0.0964 & 0.2182 & 0.0922 & 0.0393 & 0.0799 & 0.4791 & 0.2690 & 0.4405 \\
  \bottomrule
  \end{tabular}
  }
  \label{tab:rouge_dnc}
\end{table*}

\section{Conclusions}
We present \model, a multimodal chat model for RNA sequences that enhances LLM-based question-answering and accelerates RNA discovery by providing concise, accurate responses to complex queries. \model aligns RNA embeddings from the RNA-FM encoder with natural language in LLMs like Llama-3 using a learnable projection layer. We optimize instruction tuning with GPT-4o-mini to ensure high-quality, precise answers. We also introduce \dataset, a \datasize RNA question-answering dataset derived from the extensive RNA research literature. Our scalable framework, which uses topic modeling and divide-and-conquer summarization, enables efficient RNA-to-language dataset curation.

Experiments with our initial model trained on a 5K subset of \dataset show that \model generates very promising responses already, achieving high semantic and lexical scores on unseen RNA data. The full \dataset provides even more value in the development of multimodal RNA LLMs. \model with \dataset will inspire further innovations in LLM-based RNA research, driving rapid advancements in the field.

\newpage
\bibliographystyle{unsrt}
\bibliography{main}

\appendix 
\label{sec:appendix}
\section{Training Details}

We conducted initial training of \model on a 5K subset of the \dataset dataset to generate the initial model checkpoints.

During the modality alignment (\textbf{\textit{MA}}) stage (Stage 1), we optimized the projection layer over 10 epochs using a batch size of 1, weight decay of 0.05, and 2,048 warm-up steps. The AdamW optimizer~\cite{loshchilov2019decoupled} was employed with $\beta_{1} = 0.9$ and $\beta_{2} = 0.98$. A learning rate scheduler with a linear warm-up followed by cosine annealing was applied. Specifically, the initial learning rate was set to $1 \times 10^{-4}$, the minimum learning rate to $8 \times 10^{-5}$, and the warm-up learning rate to $1 \times 10^{-6}$. Automatic Mixed Precision (AMP)~\cite{micikevicius2018mixed} was employed to enhance training efficiency.

For the instruction tuning (\textbf{\textit{IT}}) stage (Stage 2), the \model architecture was further trained for an additional 10 epochs, maintaining the same batch size of 1, weight decay of 0.05, and reducing the warm-up steps to 200. The configuration of the AdamW optimizer and the use of AMP remained consistent with those used in Stage 1.

The training process was executed on two Nvidia A100 GPUs (40GB) and required a total of 1 week for both stages combined.

We will extend the training process by scaling \model to a larger configuration using the full \datasize dataset from \dataset.

\section{Case Studies}

\begin{figure}[htbp]
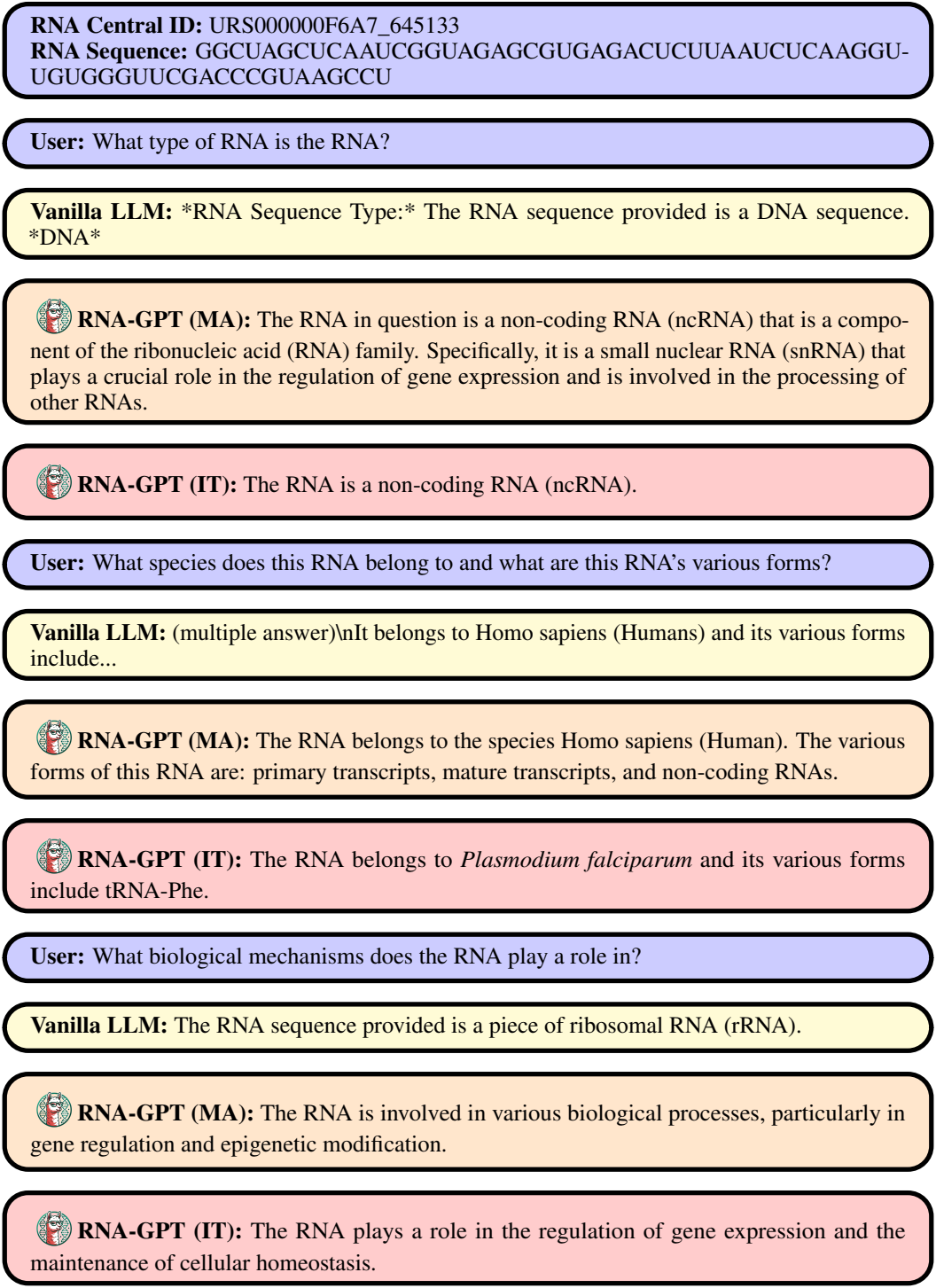

\begin{mdframed}[backgroundcolor=blue!20] 
\textbf{RNA Central ID:} URS000000F6A7\_645133 \\
\textbf{RNA Sequence:} GGCUAGCUCAAUCGGUAGAGCGUGAGACUCUUAAUCUCAAGGUUGUGGGUUCGACCCGUAAGCCU  
\end{mdframed}

\begin{mdframed}[backgroundcolor=blue!20] 
\textbf{User:} What type of RNA is the RNA?
\end{mdframed}

\begin{mdframed}[backgroundcolor=yellow!20] 
\textbf{Vanilla LLM:} *RNA Sequence Type:* The RNA sequence provided is a DNA sequence. *DNA*
\end{mdframed}

\begin{mdframed}[backgroundcolor=orange!20] 
\textbf{\raisebox{-1.25ex}{\includegraphics[height=20pt]{figures/RNA_GPT.pdf}}RNA-GPT (MA):} The RNA in question is a non-coding RNA (ncRNA) that is a component of the ribonucleic acid (RNA) family. Specifically, it is a small nuclear RNA (snRNA) that plays a crucial role in the regulation of gene expression and is involved in the processing of other RNAs.
\end{mdframed}

\begin{mdframed}[backgroundcolor=red!20] 
\textbf{\raisebox{-1.25ex}{\includegraphics[height=20pt]{figures/RNA_GPT.pdf}}RNA-GPT (IT):} The RNA is a non-coding RNA (ncRNA).
\end{mdframed}

\begin{mdframed}[backgroundcolor=blue!20] 
\textbf{User:} What species does this RNA belong to and what are this RNA's various forms?
\end{mdframed}

\begin{mdframed}[backgroundcolor=yellow!20] 
\textbf{Vanilla LLM:} (multiple answer)\textbackslash
nIt belongs to Homo sapiens (Humans) and its various forms include...
\end{mdframed}

\begin{mdframed}[backgroundcolor=orange!20] 
\textbf{\raisebox{-1.25ex}{\includegraphics[height=20pt]{figures/RNA_GPT.pdf}}RNA-GPT (MA):} The RNA belongs to the species Homo sapiens (Human). The various forms of this RNA are: primary transcripts, mature transcripts, and non-coding RNAs.
\end{mdframed}

\begin{mdframed}[backgroundcolor=red!20] 
\textbf{\raisebox{-1.25ex}{\includegraphics[height=20pt]{figures/RNA_GPT.pdf}}RNA-GPT (IT):} The RNA belongs to \emph{Plasmodium falciparum} and its various forms include tRNA-Phe.
\end{mdframed}

\begin{mdframed}[backgroundcolor=blue!20] 
\textbf{User:} What biological mechanisms does the RNA play a role in?
\end{mdframed}

\begin{mdframed}[backgroundcolor=yellow!20] 
\textbf{Vanilla LLM:} The RNA sequence provided is a piece of ribosomal RNA (rRNA).
\end{mdframed}

\begin{mdframed}[backgroundcolor=orange!20] 
\textbf{\raisebox{-1.25ex}{\includegraphics[height=20pt]{figures/RNA_GPT.pdf}}RNA-GPT (MA):} The RNA is involved in various biological processes, particularly in gene regulation and epigenetic modification.
\end{mdframed}

\begin{mdframed}[backgroundcolor=red!20] 
\textbf{\raisebox{-1.25ex}{\includegraphics[height=20pt]{figures/RNA_GPT.pdf}}RNA-GPT (IT):} The RNA plays a role in the regulation of gene expression and the maintenance of cellular homeostasis.
\end{mdframed}

\caption{Interaction between user and RNA-GPT on  URS000000F6A7\_645133}
\end{figure}

\begin{figure}[htbp]
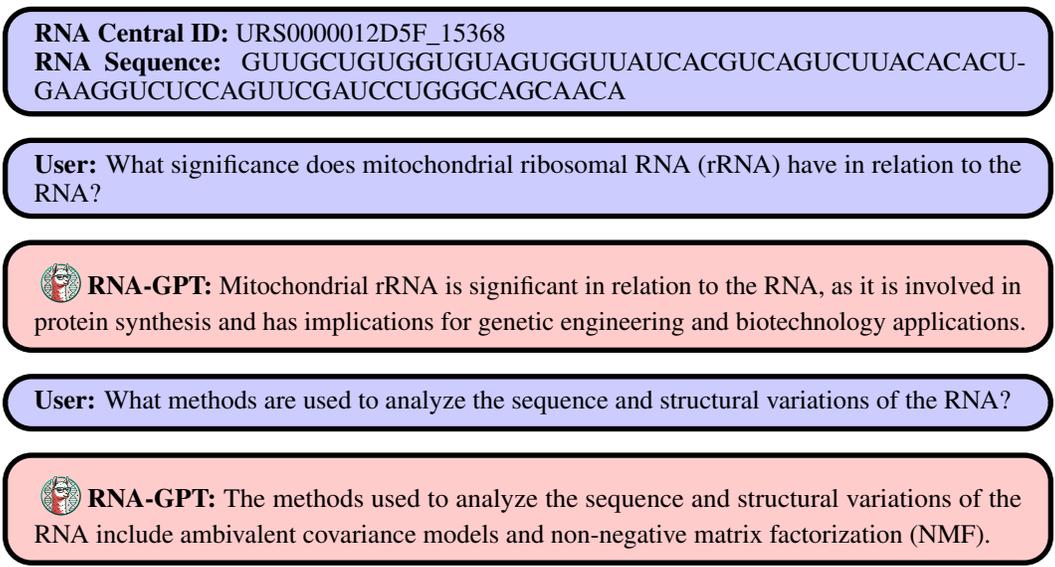

\begin{mdframed}[backgroundcolor=blue!20] 
\textbf{RNA Central ID:} URS0000012D5F\_15368 \\
\textbf{RNA Sequence:} GUUGCUGUGGUGUAGUGGUUAUCACGUCAGUCUUACACACUGAAGGUCUCCAGUUCGAUCCUGGGCAGCAACA
\end{mdframed}

\begin{mdframed}[backgroundcolor=blue!20] 
\textbf{User:} What significance does mitochondrial ribosomal RNA (rRNA) have in relation to the RNA?
\end{mdframed}

\begin{mdframed}[backgroundcolor=red!20] 
\textbf{\raisebox{-1.25ex}{\includegraphics[height=20pt]{figures/RNA_GPT.pdf}}RNA-GPT:} Mitochondrial rRNA is significant in relation to the RNA, as it is involved in protein synthesis and has implications for genetic engineering and biotechnology applications.
\end{mdframed}

\begin{mdframed}[backgroundcolor=blue!20] 
\textbf{User:} What methods are used to analyze the sequence and structural variations of the RNA?
\end{mdframed}

\begin{mdframed}[backgroundcolor=red!20] 
\textbf{\raisebox{-1.25ex}{\includegraphics[height=20pt]{figures/RNA_GPT.pdf}}RNA-GPT:} The methods used to analyze the sequence and structural variations of the RNA include ambivalent covariance models and non-negative matrix factorization (NMF).
\end{mdframed}

\caption{Interaction between user and RNA-GPT on  URS0000012D5F\_15368}
\end{figure}

\begin{figure}[htbp]
\begin{mdframed}[backgroundcolor=blue!20] 
\textbf{RNA Central ID:} URS000006DA05\_9925 \\
\textbf{RNA Sequence:} UGAAGUGCUGUGGAUUUCUUUGUGAAUCACCAUAUCUAAGCUAAUGUGGUGGUGGUUUACAAAGUAAUUCAUAGUGCUUCA
\end{mdframed}

\begin{mdframed}[backgroundcolor=blue!20] 
\textbf{User:} What diseases are associated with the RNA?
\end{mdframed}

\begin{mdframed}[backgroundcolor=red!20] 
\textbf{\raisebox{-1.25ex}{\includegraphics[height=20pt]{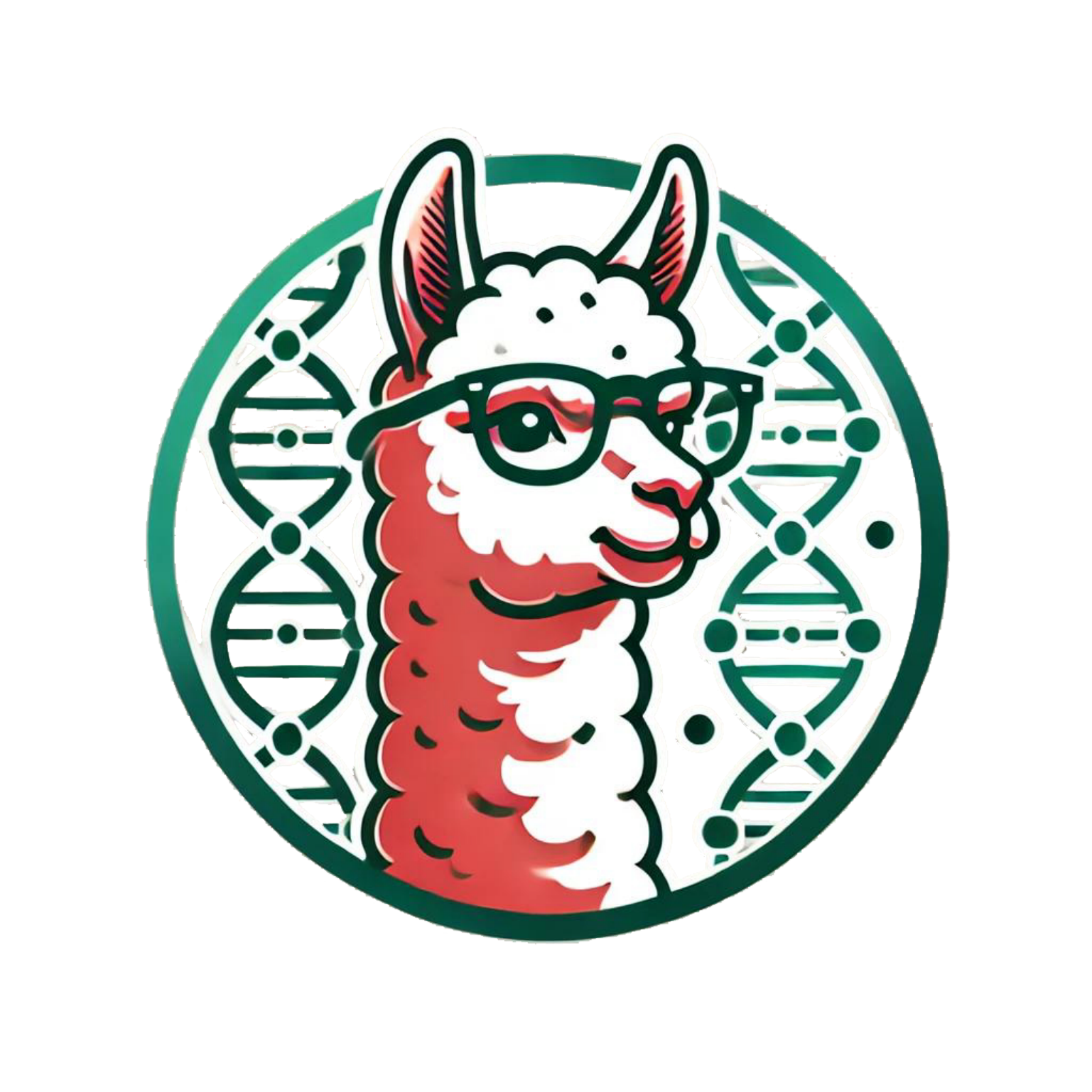}}RNA-GPT:} The RNA is associated with osteosarcoma, breast cancer, and epilepsy.
\end{mdframed}

\begin{mdframed}[backgroundcolor=blue!20] 
\textbf{User:} How is the RNA implicated in cholangiocarcinoma (CCA)?
\end{mdframed}

\begin{mdframed}[backgroundcolor=red!20] 
\textbf{\raisebox{-1.25ex}{\includegraphics[height=20pt]{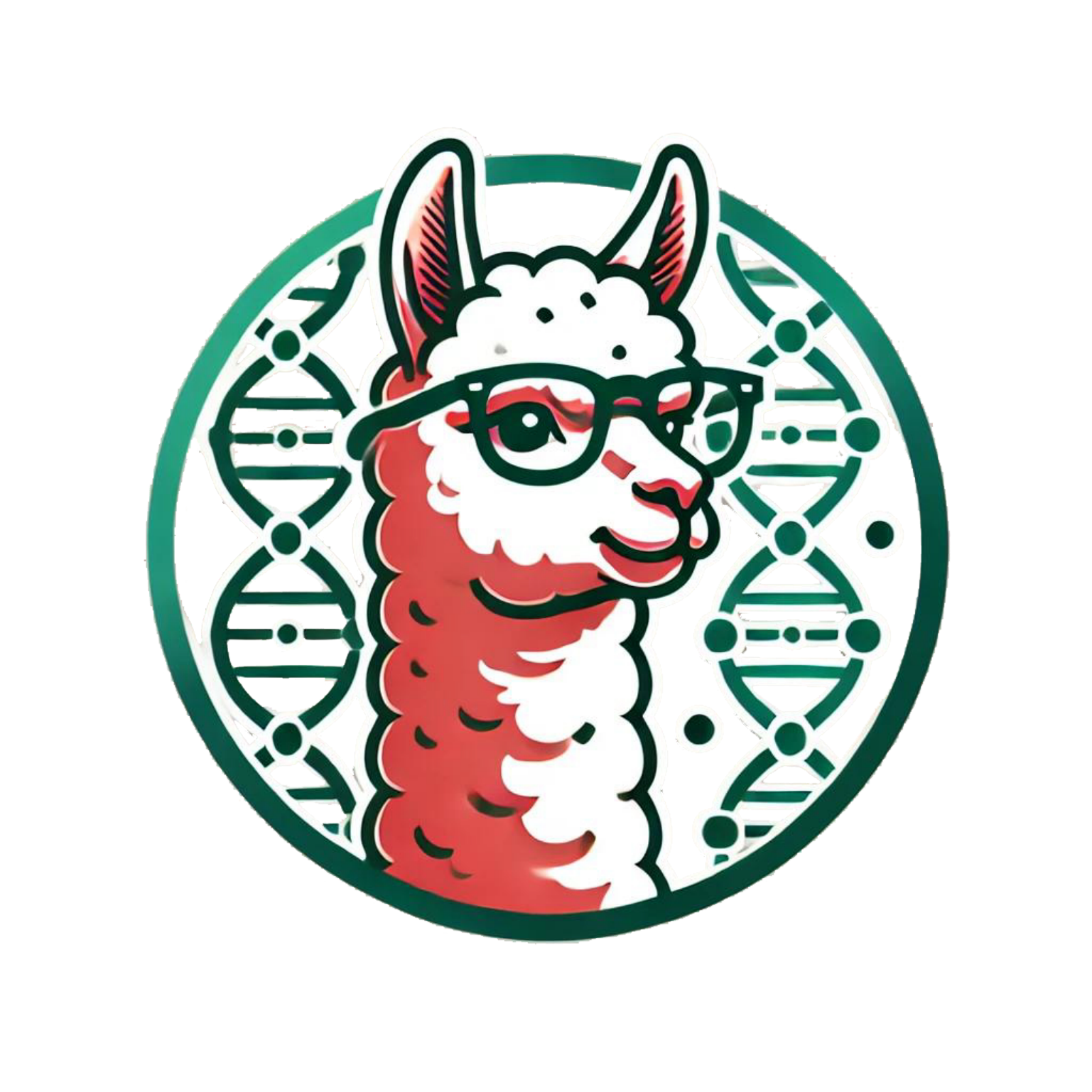}}RNA-GPT:} The RNA is significantly upregulated in CCA and may serve as a biomarker for early detection and diagnosis.
\end{mdframed}

\begin{mdframed}[backgroundcolor=blue!20] 
\textbf{User:} What is the role of the RNA in CCA cell growth?
\end{mdframed}

\begin{mdframed}[backgroundcolor=red!20] 
\textbf{\raisebox{-1.25ex}{\includegraphics[height=20pt]{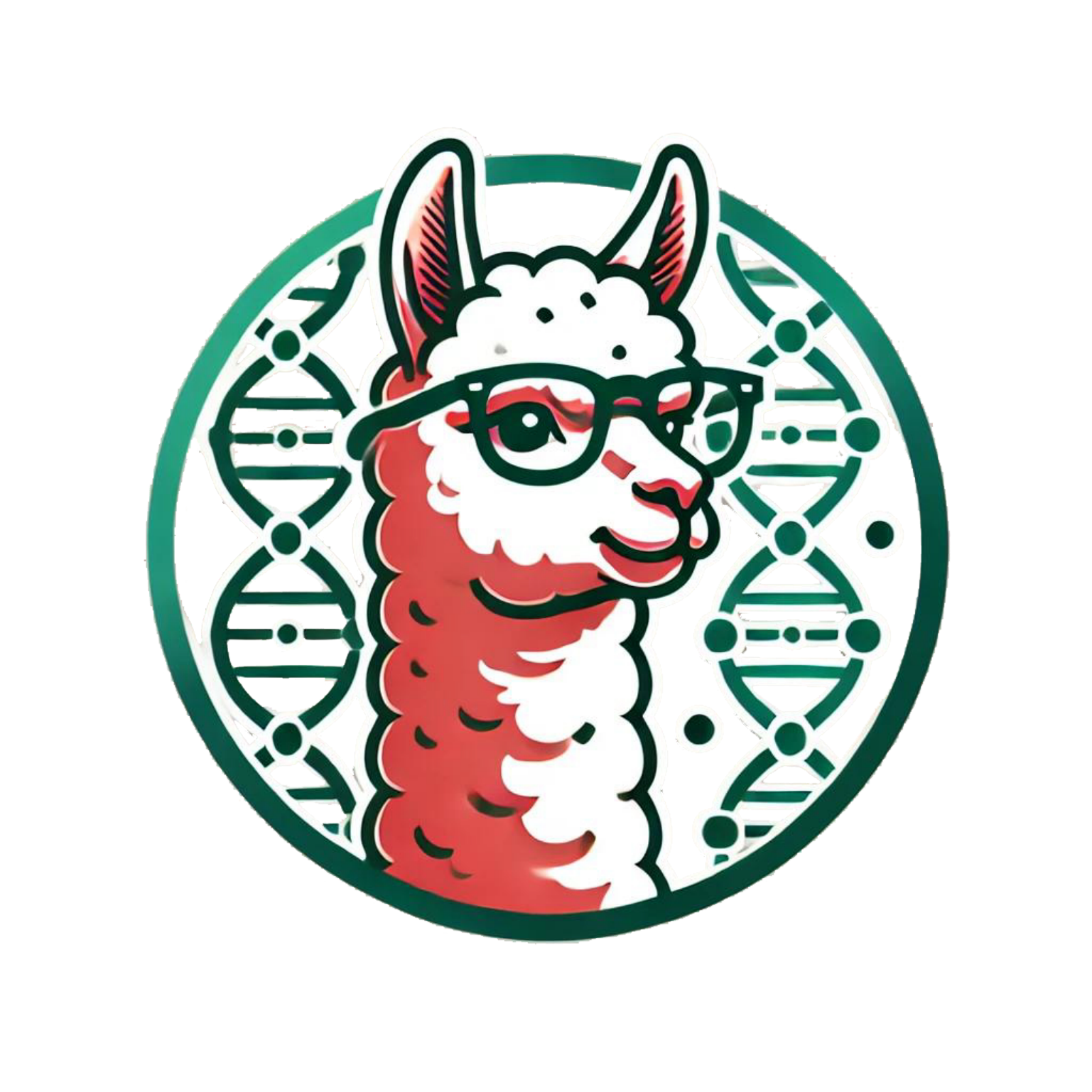}}RNA-GPT:} In CCA cell growth, the RNA acts as a tumor suppressor by regulating BCL-XL expression.
\end{mdframed}

\begin{mdframed}[backgroundcolor=blue!20] 
\textbf{User:} What is the role of circular RNAs (circRNAs) in relation to this RNA?
\end{mdframed}

\begin{mdframed}[backgroundcolor=red!20] 
\textbf{\raisebox{-1.25ex}{\includegraphics[height=20pt]{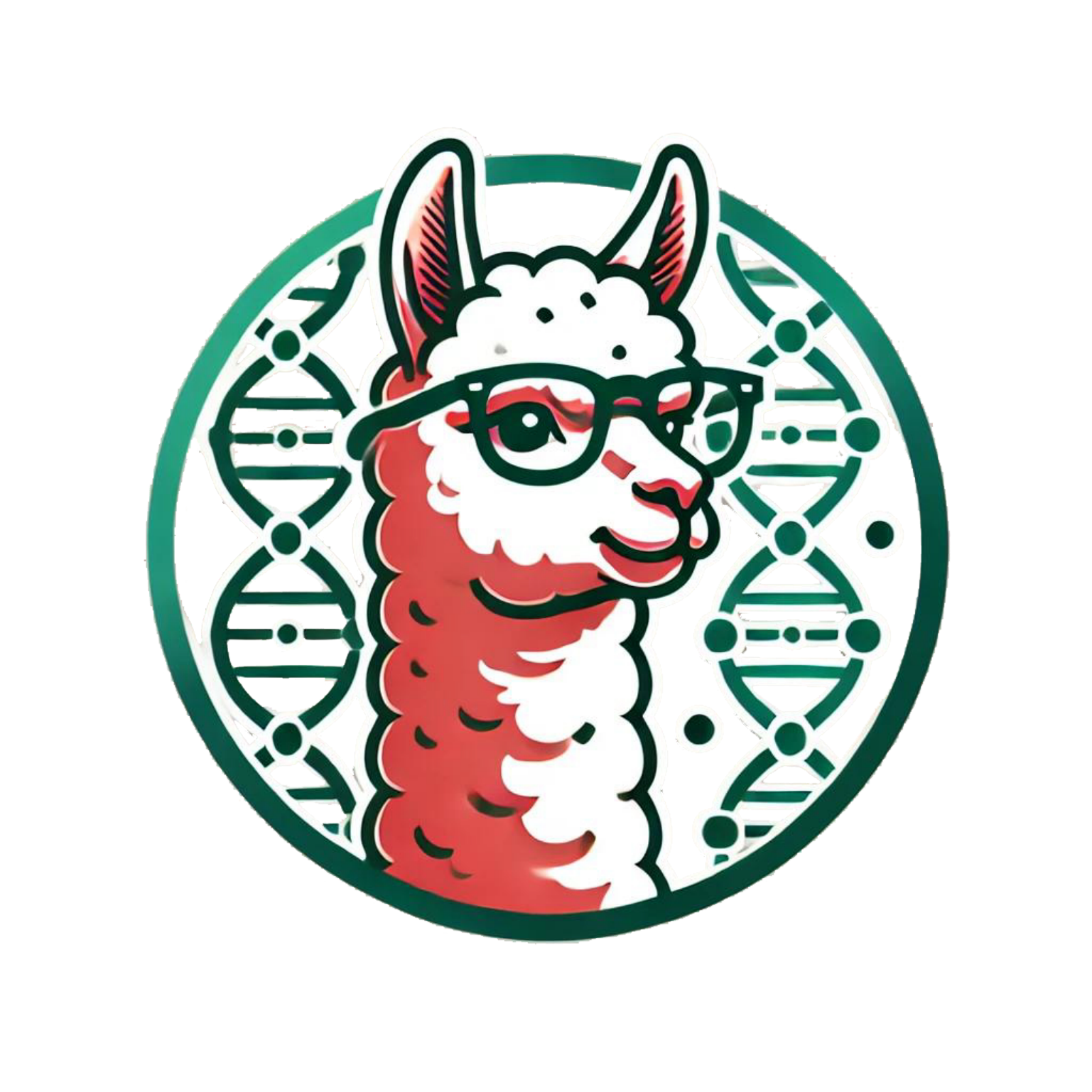}}RNA-GPT:} CircRNAs play a significant role in the regulation of gene expression and have been linked to various diseases, including cancer.
\end{mdframed}

\caption{Interaction between user and RNA-GPT on  URS000006DA05\_9925 (Part 1)}
\end{figure}

\begin{figure}[htbp]

\begin{mdframed}[backgroundcolor=blue!20] 
\textbf{User:} How does the knockdown of CDR1as affect gastric cancer (GC) cells?
\end{mdframed}

\begin{mdframed}[backgroundcolor=red!20] 
\textbf{\raisebox{-1.25ex}{\includegraphics[height=20pt]{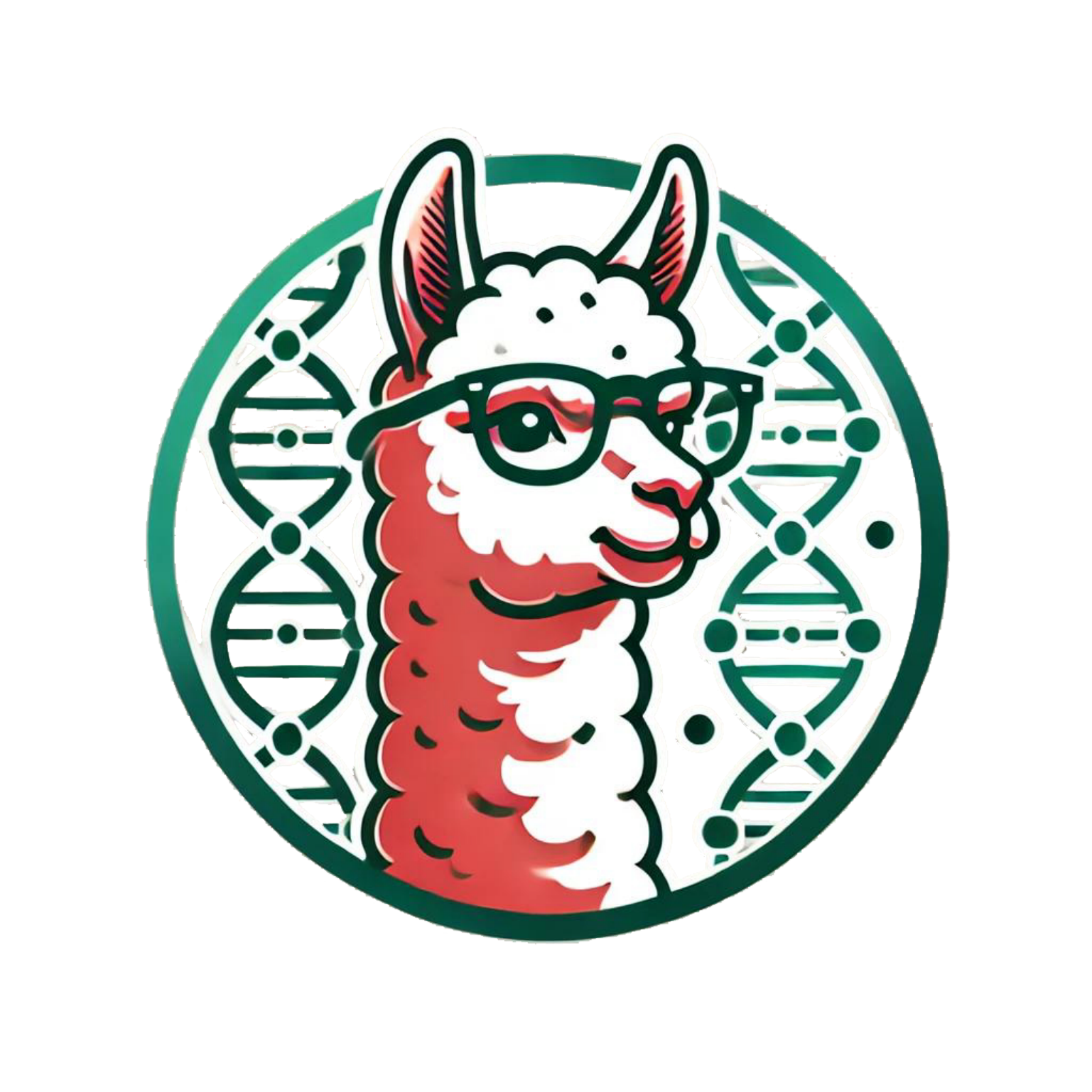}}RNA-GPT:} The knockdown of CDR1as in GC cells inhibits cell proliferation, migration, and invasion, indicating its potential as a therapeutic target.
\end{mdframed}

\begin{mdframed}[backgroundcolor=blue!20] 
\textbf{User:} How does the RNA function as a tumor suppressor in CCA treatment?
\end{mdframed}

\begin{mdframed}[backgroundcolor=red!20] 
\textbf{\raisebox{-1.25ex}{\includegraphics[height=20pt]{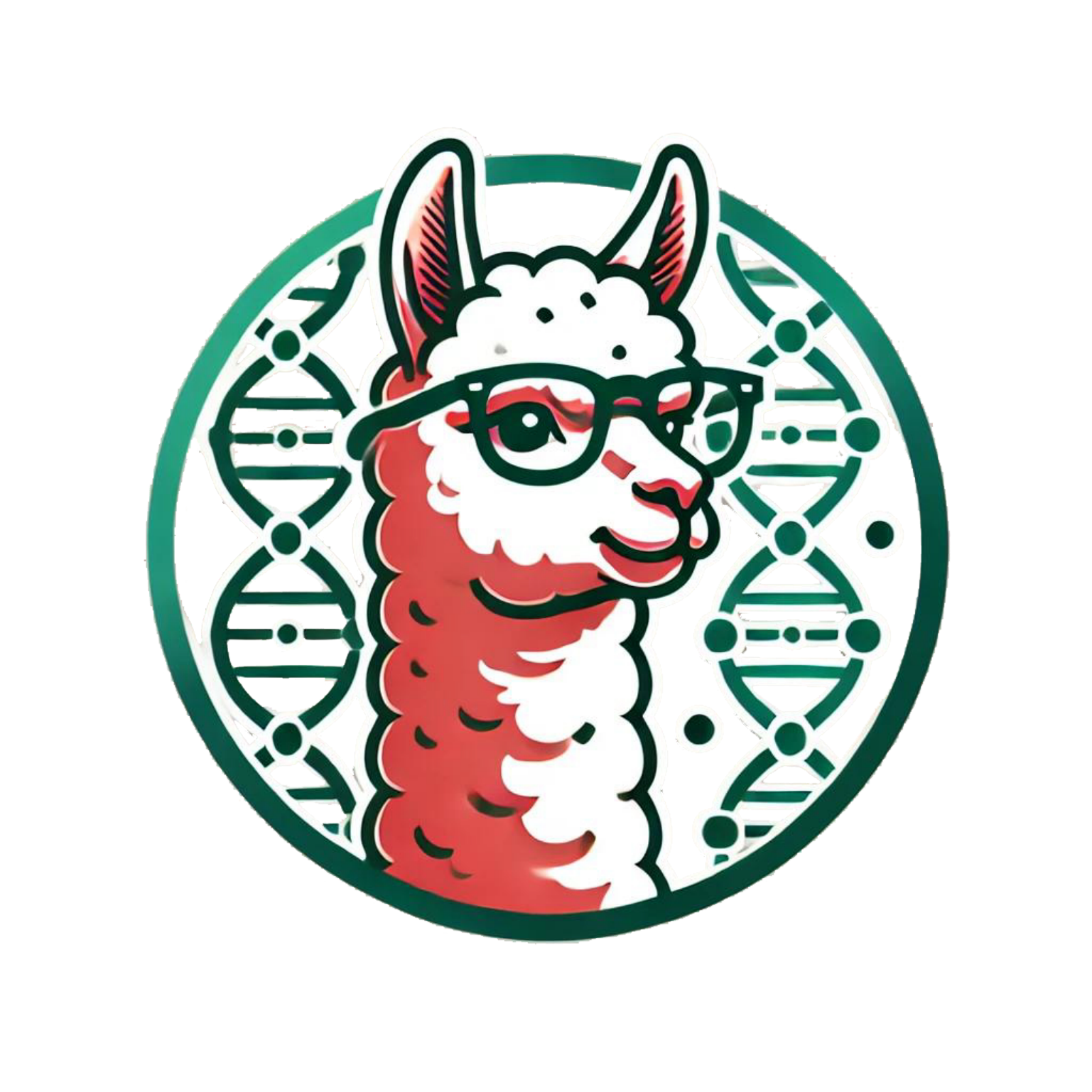}}RNA-GPT:} The RNA functions as a tumor suppressor by regulating BCL-XL and BCL-2 expression, which are key regulators of apoptosis.
\end{mdframed}

\begin{mdframed}[backgroundcolor=blue!20] 
\textbf{User:} What impact does the RNA have on myotube area and fusion index in DM1?
\end{mdframed}

\begin{mdframed}[backgroundcolor=red!20] 
\textbf{\raisebox{-1.25ex}{\includegraphics[height=20pt]{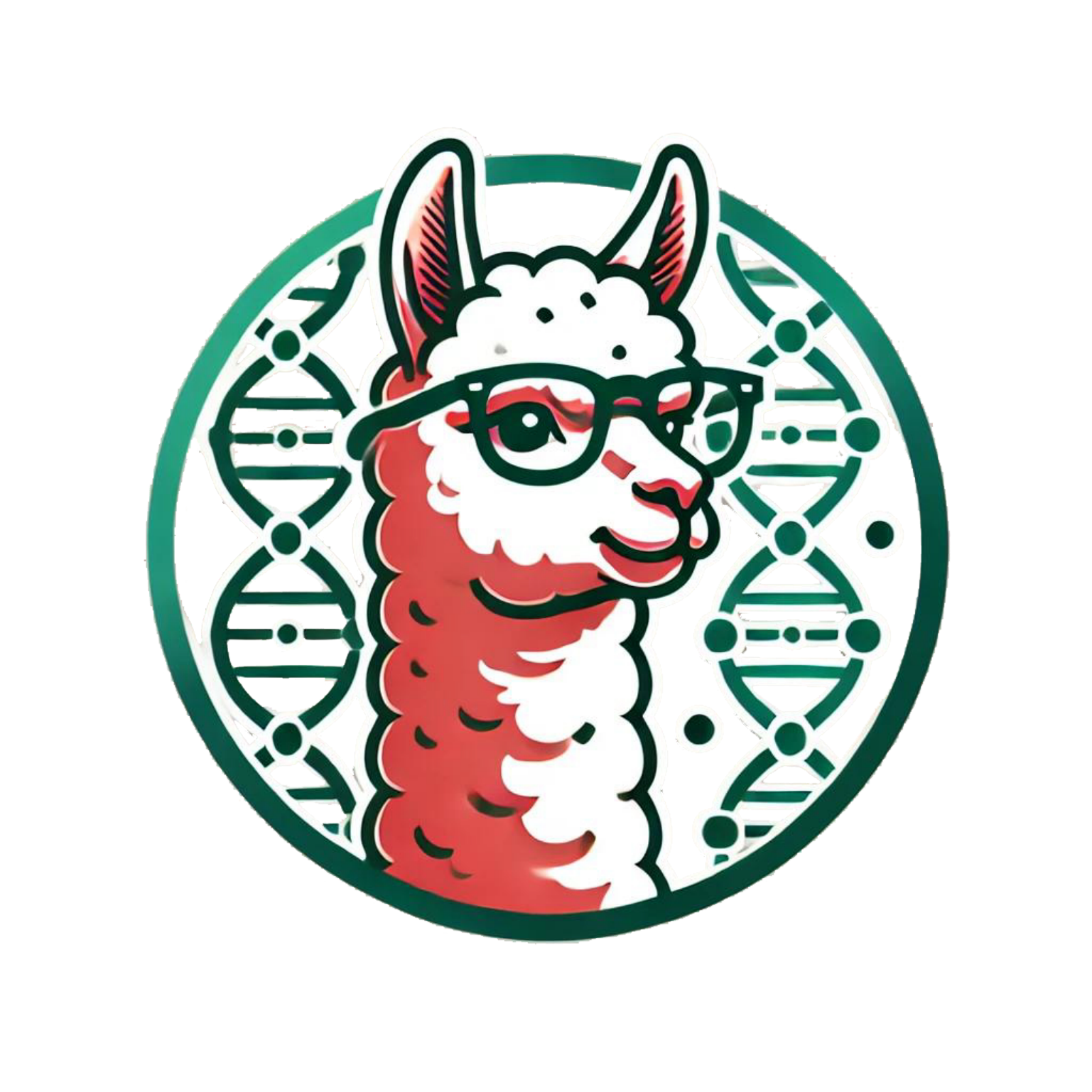}}RNA-GPT:} The RNA has a significant impact on myotube area and fusion index in DM1, indicating its potential role in muscle regeneration.
\end{mdframed}

\begin{mdframed}[backgroundcolor=blue!20] 
\textbf{User:} What species does this RNA belong to and what are this RNA's various forms?
\end{mdframed}

\begin{mdframed}[backgroundcolor=red!20] 
\textbf{\raisebox{-1.25ex}{\includegraphics[height=20pt]{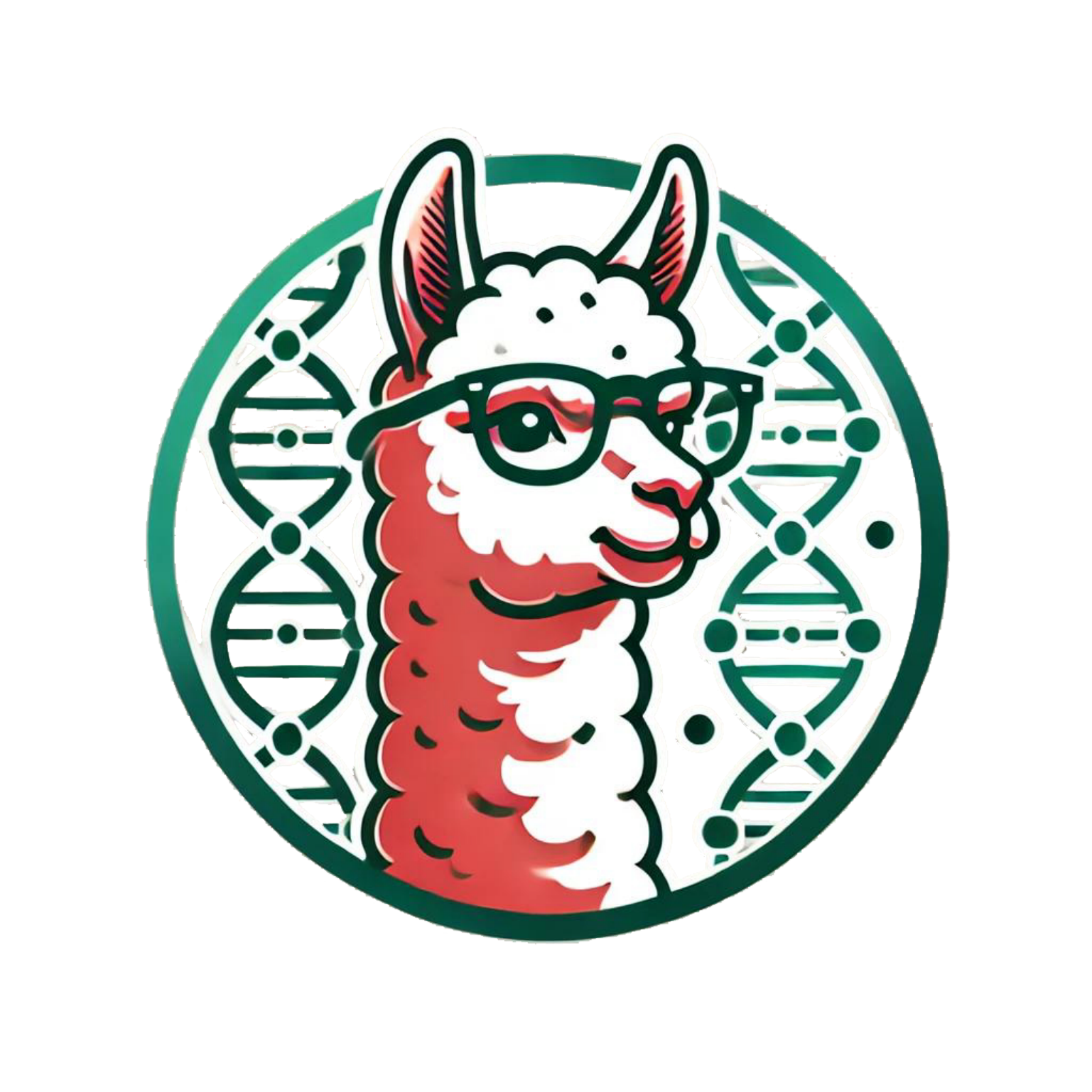}}RNA-GPT:} The RNA belongs to the chimpanzee (\emph{Pan troglodytes}) and its various forms include miRNAs, long non-coding RNAs, and circular RNAs.
\end{mdframed}

\caption{Interaction between user and RNA-GPT on  URS000006DA05\_9925 (Part 2)}
\end{figure}

\end{document}